\newcommand{\la}{\langle}
\newcommand{\ra}{\rangle}
\newcommand{\beq}{\begin{eqnarray}}
\newcommand{\eeq}{\end{eqnarray}}

\newcommand{\btem}{\bibitem}

\documentstyle[12pt]{article}    

\begin{document}

\setlength{\baselineskip}{0.3in}

\rightline{UTHEP-337}

\bigskip

\begin{center}

{\large {\bf Tensor Charge of the Nucleon on the Lattice}
\footnote{Invited talks  presented at
 the  Australia-Japan
 workshop on Quarks, Hadrons and Nuclei (Nov. 15-24, 1995, Adelaide, 
 Australia) and at RIKEN International Symposium on
 Spin Structure of the Nucleon, (Dec. 18-19, 1995, RIKEN, Japan)} }\\

\vspace{0.2cm}
 
  S. Aoki, M. Doui and T. Hatsuda  \\

\bigskip
 Institute  of Physics,  University of Tsukuba, Tsukuba, Ibaraki 305, Japan\\

\end{center}

\vspace{1.2cm}

\centerline{abstract}

 Tensor charge of the nucleon, which will be 
  measured in  Drell-Yan
 processes in polarized proton-proton  collisions at RHIC, are
 studied in quenched lattice QCD simulation.
 On the 16$^3\times$ 20 lattice with $\beta=5.7$, 
 connected parts of the tensor charge are determined with small
 statistical error, while the disconnected parts are found to be 
 small with relatively large error bars.  Flavor-singlet tensor
 charge ($\delta \Sigma = \delta u + \delta d + \delta s$)
 is not suppressed as opposed to the flavor singlet 
 axial charge ($\Delta \Sigma = \Delta u + \Delta d + \Delta s$).

\vspace{2cm}
\setcounter{section}{0}

\section{Introduction}

  The parton structure of the nucleon in the twist 2 level is known to be
 characterized by three structure functions 
 $f_1(x,\mu)$, $g_1(x,\mu)$ and $h_1(x,\mu)$
 with $x$ being the Bjorken variable and $\mu$ being the renormalization
 scale (see e.g., \cite{JJ}). 
 $f_1$ and $g_1$, which represent the quark-momentum distribution 
 and quark-spin distribution respectively, 
 can be measured by the deep inelastic lepton-hadron
 scattering (DIS).  On the other hand, $h_1$, which represents  
 the quark-transversity distribution, can only be measured in the
 polarized Drell-Yan processes, since it is related to the 
 matrix element of the chiral-odd quark operator.
 Although such experiments are not yet available, 
 it is planned in RHIC at BNL. Therefore, theoretical prediction
 on $h_1$ has some importance. Also, whether there is a 
  ``transversity crisis'' in the first moment of $h_1$ as in the case
 of ``spin crisis'' in $g_1$ is an interesting question to be 
 examined. 
 In this paper, we will concentrate on the 
 first moment of $h_1(x)$ and report  our recent studies 
 using lattice QCD simulations \cite{ADH}.

\section{Tensor charge versus axial charge}

\subsection{definitions}

The first moments of $g_1(x,\mu)$ and $h_1(x,\mu)$ are
 called $\Delta q (\mu)$ and $\delta q(\mu)$ respectively. They
 are related to the nucleon matrix elements  of the
 axial current and tensor current as follows
\beq
\la ps \mid \bar{q} \gamma_{\mu} \gamma_5 q \mid ps \ra & = & 
 2 M s_{\mu}\  \Delta q , \label{def0} 
\\
\la ps \mid \bar{q} i\sigma_{\mu \nu} \gamma_5 q \mid ps \ra & = & 
 2 (s_{\mu} p_{\nu} - s_{\nu} p_{\mu})\  \delta q , 
\label{def1}
\eeq
where $p_{\mu}$ is the nucleon's four momentum, $M$ is the nucleon's
  rest mass,
 and $s_{\mu}$ is the nucleon's covariant spin-vector.

 In  the light cone  frame,
   $\Delta q$ is interpreted as total quark-helicity in the 
 nucleon, while $\delta q$ is a total quark-transversity in the
  nucleon \cite{JJ}:
\beq
\Delta q (\mu) & = & \int_0^1 [g_1(x,\mu) + \bar{g}_1(x,\mu)]dx \\ \nonumber
 & = & \int_0^1 [N_+(x,\mu) - N_-(x,\mu) + \bar{N}_+(x,\mu) -
 \bar{N}_-(x,\mu)] dx , 
\eeq
and
\beq
\delta q (\mu) & = & \int_0^1 [h_1(x,\mu) - \bar{h}_1(x,\mu)]dx \\ \nonumber
 & = & \int_0^1 [N_{\uparrow}(x,\mu) - N_{\downarrow}(x,\mu) -
 \bar{N}_{\uparrow}(x,\mu) + \bar{N}_{\downarrow}(x,\mu)] dx . 
\eeq
Here $N_+(x)$ ($N_-(x)$) denotes
 the momentum distribution of quarks having the same (opposite) helicity with
 the nucleon, while  
  $N_{\uparrow}(x)$ ($N_{\downarrow}(x)$) denotes
 that having the same (opposite) transverse polarization
  with the nucleon.  The quantity with bar is the
 distribution for anti-quark.

On the other hand, in the rest frame of the nucleon,
 $\Delta q$ ($\delta q$)
 denotes the quark-spin + anti-quark-spin 
 (quark-spin $-$ anti-quark-spin), which can be seen by taking
 $\mu = i$ $(\mu =0, \nu =i)$ in eq. (\ref{def0},\ref{def1}).
 In this frame, estimates by using hadron models are possible.
 For example, relativistic quark models which can reproduce 
 $g_A = 1.25$ correctly give simple inequalities;
\beq
\mid \delta u \mid > \mid \Delta u \mid \ ,
 \ \ \ \ \ \ \ 
\mid \delta d \mid > \mid \Delta d \mid \ .
\label{bag}
\eeq
The lower component of the Dirac spinor
 of the confined quarks plays essential role for the above inequalities.

 Drawbacks  of such model-calculations are  (i) 
 the renormalization scale where the matrix elements are 
 evaluated is not clear, and (ii) strange quark contribution, which 
 originates from the OZI violating processes, is hard
 to estimate in a reliable manner.
 Lattice QCD simulations can overcome these problems 
 even within the quenched approximation.  
 In particular, $\Delta u$, $\Delta d$ and $\Delta s$
 have been studied by two groups \cite{kek,liu} and their results
 are consistent with the recent experimental data on the 
 spin structure of the nucleon \cite{AR}.
  In the quenched level,  two kinds of diagrams
 arise:  One is the connected amplitude (Fig.1a)
 in which the external operator is connected to one of the
 valence nucleon lines, and another is the disconnected
 amplitude (Fig.1b) where the quark line coming from the
 external operator is closed by itself.
 The latter gives OZI violating amplitude such as the
 strangeness content in the nucleon.

\vspace{8cm}
\noindent
{\bf FIG.1:} (a) Connected matrix element of the 
 nucleon.  Cross denotes the operator insertion.
 (b) Disconnected matrix element of the nucleon.

\vspace{0.5cm}

\section{Matrix elements on the lattice}

\subsection{mass and matrix elements}

We use the standard Wilson's  action
 in our simulation in which  two basic
 parameters read  $\beta \equiv 6/g^2$
 ($g$ being the bare gauge coupling) 
 and the hopping parameter $K$. 
  In the following, instead of $K$, we use the 
 ``quark-mass'' $ma \equiv (1/K - 1/K_c)/2 $ where
 $a$ is the lattice spacing and $K_c$ is the critical 
hopping parameter at which the pion becomes massless.
  
 Hadron masses are obtained by the correlation function of 
 composite operators for large time $t$. For example, 
 the nucleon mass is obtained from 
\beq
 \la N(t) \bar{N}(0) \ra \rightarrow  {\rm const.} \times e^{-m_N t} ,
\eeq
with $N(t)$ being the spatially integrated
 interpolating operator for the nucleon 
 $N(t) = \int d^3x (q C^{-1} \gamma_5 q)q$.
   On the other hand,  matrix element of local operator is
 obtained as 
\beq
 R(t) \equiv {\la N(t) \sum_{t',x} O(t',x) \bar{N}(0) \ra \over 
\la N(t) \bar{N}(0) \ra }  \rightarrow  {\rm const.}
 + \la N \mid O \mid N \ra \ t \ .
\eeq
Namely the linear slope of $R( t \rightarrow {\rm large})$  gives
 matrix element defined on the lattice.

\subsection{some remarks}

In our actual simulation, the following points have been 
 considered  \cite{kek}.
 
\noindent
(i)  To get large overlap of $N(0)$ with the real nucleon, we
 use wall source at initial time slice $t=0$.
 To do this, we made Coulomb gauge fixing
 only at $t=0$.  
  (ii)  To avoid mirror source at the final time slice
 $t=t_f$,  we set
 Dirichlet boundary condition at $t_f$.
  (iii)  To compare the  matrix element on the lattice with that
 in the $MS$-bar scheme, we multiplied a
  renormalization factor $Z(\mu a)$ to the obtained data.
 For the tensor charge, $Z$ calculated for $\mu =1/a$ using
 tadpole improved perturbation \cite{LM} reads
\beq
 Z = \left( 1- {3 K \over 4 K_c} \right) \left[ 
 1- 0.44\  \alpha_s(1/a) \right] .
\eeq

\section{Results}

 We have done simulations in three cases:
 $12^3 \times 20$ ($\beta = 5.7$), 
 $16^3 \times 20$ ($\beta = 5.7$), 
 $16^3 \times 20$ ($\beta = 6.0$). In this 
  report, we will show the results of the second simulation.
  We have used Fujitsui VPP500 and analyzed 1053 gauge
 configurations. Each configuration is taken after every
 1000 sweeps.  Three different values of the quark masses
 $K=0.160, 0.164, 0.1665$ are adopted to extract physical
 quantities  in the chiral limit.
 The statistical errors are estimated by the jackknife procedure.

 Hadron masses are extracted by the
 $\chi^2$ fitting of the
 data in the interval $5 \le t \le 10$.
 By using the physical hadron masses, $m_{\pi, \rho, K}
  = 135, 770, 498 $ MeV, one obtains
$a^{-1} = 1.42 $GeV ($a=0.14$ fm), $m = 4.8$ MeV
 and $m_s =$ 125 MeV.  This also predict the nucleon mass
 as $m_N = 1.13 $ GeV.  The physical volume of the lattice
 is $V = (2.24)^3 \times 2.8 $ fm$^4$.

 $\chi^2$ fitting  in the interval $6 \le t \le 11$ is also
 applied for the connected and disconnected part of the
 matrix elements.
 Shown in Fig.2 are the  data
 for the correlation function $R(t)$ for $K=0.164$.
 For connected $u,d$ contributions ($\delta u_{con.},
 \delta d_{con.}$) , one can see a clear non-vanishing
 linear slope in  $5 \le t \le 10$, while disconnected $u-d$
 contribution ($\delta u_{dis.}, \delta d_{dis.})$
  has very small slope evenif it exists.

\newpage

\ \ \ 

\vspace{18cm}
\noindent
{\bf FIG.2:}  $R(t)$ as a function of $t$
 for medium-heavy quark mass $K=0.164$.
 The black (white) circle denotes the connected amplitude
 $\delta u_{con.} (\delta d_{con.})$, while
  the black triangle denotes the disconnected amplitude
    $\delta u_{dis.} = \delta d_{dis.}$.

\newpage

 Table 1 shows the result of the data extrapolated down 
 to the chiral limit.  The result is compared with that
 for $\Delta q$  in  \cite{kek} with
 the same lattice size and $\beta$.
  
\vspace{0.5cm}

\begin{center}

\begin{tabular}{|c|c|}   \hline 

Tensor charge $\delta q$ (this work)
 & Axial charge $\Delta q$   (ref.[3])  \\
$16^3 \times 20$, $\beta=5.7$  & $16^3 \times 20$, $\beta=5.7$     \\
1053 gauge configurations       &    260 gauge configurations       \\
 $(\mu^2 = 3 $ GeV$^2$)        &   $(\mu^2 = 3 $ GeV$^2$)         \\ \hline
\hline
 $\delta u_{con.} = +0.89\  (2)$  &  $\Delta u_{con.} = +0.76 \ (4)$   \\ 
 $\delta u_{dis.} = -0.05\ (6)$  &  $\Delta u_{dis.} = -0.12\  (4)$   \\ 
 $\delta u_{\ \ \ \ } = +0.84 \ (6)$  &  $\Delta u_{\ \ \ \ }
 = +0.64 \ (5)$   \\
 \hline 
 $\delta d_{con.} = -0.18\  (1)$  &  $\Delta d_{con.} = -0.23\  (2)$   \\ 
 $\delta d_{dis.} = -0.05 \  (6)$  &  $\Delta d_{dis.} = -0.12 \ (4)$   \\ 
 $\delta d_{\ \ \ \ } = -0.23 \ (6)$  &  $\Delta d_{\ \ \ \ } = -0.35 \ (5)$
   \\ 
\hline
 $\delta s_{\ \ \ \ } = -0.05 \ (10)$  &  $\Delta s_{\ \ \ \ } = -0.11 \ (3)$
   \\ \hline
 $\delta \Sigma_{\ \ \ \ } = +0.56 \ (13)$  & 
 $\Delta \Sigma_{\ \ \ \ } = +0.18 \ (10) $   \\  \hline
 \end{tabular}

\end{center}
\vspace{1cm}

\noindent
Table 1:  Comparison of the tensor and axial charges measured 
on the lattice.
 $\delta u \equiv \delta u_{con.} + \delta u_{dis.}$,
 $\delta d \equiv \delta d_{con.} + \delta d_{dis.}$,
  $\delta s \equiv  \delta s_{dis.}$, and the same definitions
 also hold for $\Delta q$ .

\section{Summary and discussions}

In our simulation, we have found the following.

\begin{enumerate}

\item
  As for the connected part of the tensor/axial charge, the following
 inequalities hold:
\beq
\mid \delta u \mid > \mid \Delta u \mid \ ,
 \ \ \ \ \ \ \ 
\mid \delta d \mid < \mid \Delta d \mid \ .
\eeq
This is different from the prediction of relativistic quark models
 which have  universal inequality eq.(\ref{bag}).

\item
  The disconnected part has still large statistical error and one cannot
 make definite conclusion from our simulation. Nevertheless, there is an
 indication that (i) the disconnected part is flavor independent i.e.,
 $\delta u_{dis.} \sim \delta d_{dis.} \sim \delta s_{dis.}$, and
 (ii) they are small but slightly negative.

\item
 Flavor singlet tensor charge $\delta \Sigma = \delta u
 + \delta d + \delta s$ is not suppressed as opposed to $\Delta \Sigma$:
\beq
\delta \Sigma(3 {\rm GeV}^2) = 0.56 \ (13) \ \ \ \leftrightarrow \ \ \ \  
\Delta \Sigma(3 {\rm GeV}^2)  = 0.18 \ (10),
\eeq
which implies that there is no ``transversity crisis'' for the tensor
 charge.

\end{enumerate}

Now, what we need to understand is the origin of the smallness
 of the disconnected part of the tensor charge.  Since the
 operator $\bar{q} \sigma_{\mu \nu} \gamma_5 q$
 ($\bar{q} \gamma_{\mu} \gamma_5 q$) is a charge conjugation
 odd (even) operator, $\delta q (\Delta q$) has
 a meaning q-spin $-$ $\bar{q}$-spin (q-spin $+$ $\bar{q}$-spin).
 This indicates that there is a large cancellation of the 
quark-spin content of the nucleon and the anti-quark-spin content of the
 nucleon.   Also, disconnected tensor charge is zero in any
 order of perturbation theory in massless QCD, which
   might have some relation to its  smallness in the non-perturbative
 regime. 
  To clarify the above issue, we are currently collecting more 
 data on $\delta q_{dis.}$ and $\Delta q_{dis.}$ simultaneously 
 with  $16^3 \times 20$ lattice  $(\beta = 5.7)$.

\vspace{3cm}

\centerline{\bf Acknowledgements}

 This work was supported in part by 
  the Grants-in-Aid of the Japanese Ministry of 
Education, Science and Culture (No. 06102004).

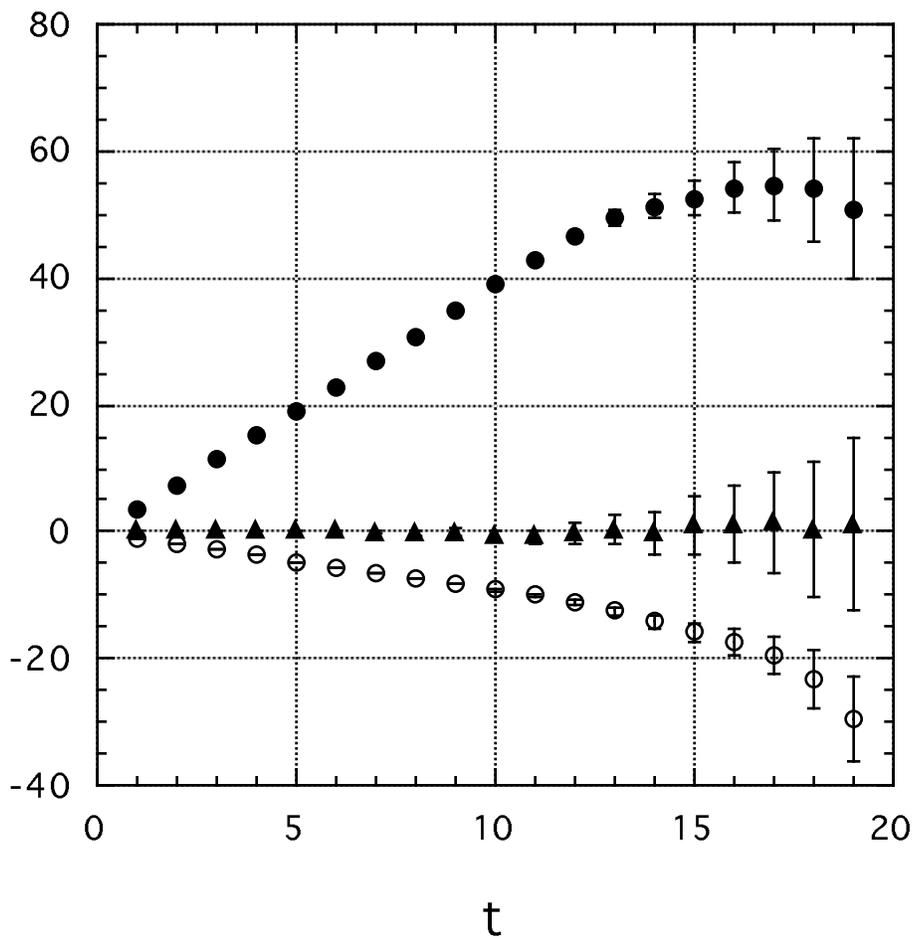
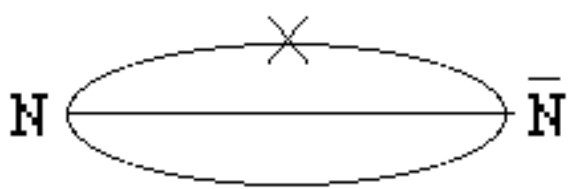
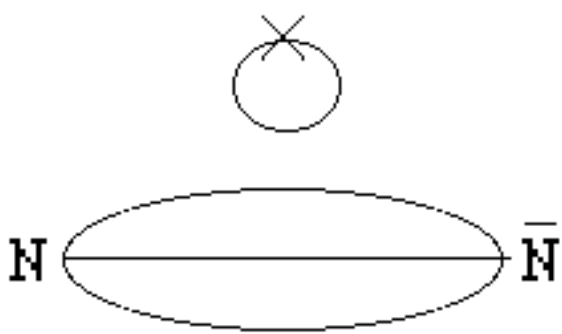
\end{document}